\documentclass[twoside]{article}

\usepackage{qic}
\usepackage{amsmath}

\newcommand{\bfF}{\mathbf{F}}

\newcommand{\hatbfx}{\hat{\mathbf{x}}}
\newcommand{\hatbfy}{\hat{\mathbf{y}}}
\newcommand{\hatbfz}{\hat{\mathbf{z}}}
\newcommand{\bfsig}{\mbox{\boldmath$\sigma$}}

\begin{document}
\setlength{\textheight}{8.0truein}  


\normalsize\textlineskip
\thispagestyle{empty}
\setcounter{page}{1}


\vspace*{0.88truein}

\alphfootnote

\fpage{1}

\centerline{\bf
HIGH FIDELITY SINGLE-QUBIT GATES USING NON-ADIABATIC RAPID PASSAGE}
\vspace*{0.37truein}
\centerline{\footnotesize
Ran Li}
\vspace*{0.015truein}
\centerline{\footnotesize\it Department of Physics, Southern Illinois
University}
\baselineskip=10pt
\centerline{\footnotesize\it Carbondale, IL 62901-4401}
\vspace*{10pt}
\centerline{\footnotesize
Melique Hoover}
\vspace*{0.015truein}
\centerline{\footnotesize\it Department of Physics, Southern Illinois 
University}
\baselineskip=10pt
\centerline{\footnotesize\it Carbondale, IL 62901-4401}
\vspace*{10pt}
\centerline{\footnotesize
Frank Gaitan\footnote{To whom correspondence should be sent.}}
\vspace*{0.015truein}
\centerline{\footnotesize\it Department of Physics, Southern Illinois
University}
\baselineskip=10pt
\centerline{\footnotesize\it Carbondale, IL 62901-4401}
\vspace*{0.225truein}

\vspace*{0.21truein}

\abstracts{
Numerical simulation results are presented which suggest that a class of 
non-adiabatic rapid passage sweeps first realized experimentally in 1991 
should be capable of implementing a set of quantum gates that is universal
for one-qubit unitary operations and whose elements operate with error 
probabilities $P_{e}<10^{-4}$. The sweeps are non-composite and 
generate controllable quantum interference effects which allow the one-qubit 
gates produced to operate non-adiabatically while maintaining high accuracy. 
The simulations suggest that the one-qubit gates produced by these sweeps show 
promise as possible elements of a fault-tolerant scheme for quantum computing. 
}{}{}

\vspace*{10pt}

\vspace*{3pt}

\vspace*{1pt}\textlineskip
\vspace*{-0.5pt}
\noindent
\section{\label{sec1}Introduction}
During the years 1997-1998 a number of researchers \cite{ft1}--\cite{ft7}
showed that under appropriate circumstances a quantum computation of arbitrary
duration could be carried out with arbitrarily small error probability in
the presence of noise and imperfect quantum logic gates. The conditions
that underlie this remarkable result are that: (1) computational data is 
protected by a sufficiently layered concatenated quantum error correcting 
code; (2) fault-tolerant protocols for quantum computation are used; and 
(3) all quantum gates used in the computation have error 
probabilities\footnote{In this paper all gate error probabilities are 
per-operation.} $\, P_{e}$ that fall below a value known as the accuracy 
threshold $P_{a}$. One of the central challenges facing the field of quantum 
computing is determining how to implement quantum gates with error 
probabilities satisfying $P_{e}<P_{a}$. The accuracy threshold has been 
calculated for a number of simple noise models yielding results in the range 
$10^{-6} < P_{a} < 10^{-3}$. For many $P_{a}\sim 10^{-4}$ has become a 
rough-and-ready working estimate for the threshold so that gates are 
anticipated to be approaching the accuracies needed for fault-tolerant quantum 
computing when $P_{e}<10^{-4}$. A number of universal sets of quantum gates 
have been found \cite{us1}--\cite{us5} and so the problem of producing 
sufficiently accurate quantum gates has shifted to producing a sufficiently 
accurate universal set of such gates. One well-known universal set consists of 
the single-qubit Hadamard, phase, and $\pi /8$ gates together with the 
two-qubit controlled-NOT gate \cite{boy}. The single-qubit gates in this set 
are sufficient to construct any single-qubit unitary operation. 

In this paper numerical simulation results are presented which suggest that 
an existing class of non-adiabatic rapid passage sweeps \cite{zw1} should be 
capable of implementing a set of quantum gates $\mathcal{S}_{1}$ that is 
universal for one-qubit unitary operations. The one-qubit gates in 
$\mathcal{S}_{1}$ are the Hadamard, NOT, $V_{p}$, and $V_{\pi /8}$ gates. The 
universality of $\mathcal{S}_{1}$ for one-qubit unitary operations is 
established by noting that the Hadamard gate is an element, and demonstrating 
that the phase and $\pi /8$ gates can be constructed from the $V_{p}$, $V_{\pi 
/8}$, and NOT gates. This is done in Section~\ref{sec3}. For each of the gates 
in $\mathcal{S}_{1}$, sweep parameter values are presented which simulations 
indicate will yield gates that operate non-adiabatically and with error 
probabilities $P_{e}< 10^{-4}$. This level of accuracy is a consequence of 
controllable quantum interference effects that are generated by these 
sweeps \cite{fg1}. We explain the optimization procedure used to search for 
sweep parameter values that (when successful) yield this high degree of gate 
accuracy. 

The outline of this paper is as follows. In the following Section we
summarize the necessary background associated with this class of non-adiabatic
rapid passage sweeps; Section~\ref{sec3} presents our simulation results for
the different gates; and Section~\ref{sec4} discusses these results, their
relation to existing work in the literature, and also current challenges.

\section{Twisted Rapid Passage}
\label{sec2}
\noindent
We consider a qubit that couples to an external control field $\bfF (t)$
through the Zeeman interaction:
\begin{equation}
\label{nonrotHam}
H(t) = \bfsig\cdot\bfF (t) \hspace{0.1in} ,
\end{equation} 
where $\bfsig$ are the Pauli matrices. The sweeps we will be interested in are 
a generalization of those used in adiabatic rapid passage (ARP). In ARP the 
field $\bfF (t)$ in the detector frame \cite{sut} is inverted over a time 
$T_{0}$ such that $\bfF (t) = b\,\hatbfx +at\,\hatbfz$. In an NMR realization 
of ARP, as seen in the lab frame, the detector frame rotates about the static 
magnetic field $B_{0}\,\hatbfz$. In the detector frame, $\hatbfz$ is chosen
to be parallel to the rotation axis. In the rotating wave approximation the 
rf-magnetic field $\mathbf{B}_{rf}$ in the lab frame lies in the $x$-$y$ plane 
and rotates about the static magnetic field. The detector frame is chosen to 
rotate with $\mathbf{B}_{rf}$ so that in this frame the rf field is static and
its direction defines $\hatbfx$: $\mathbf{B}_{rf}=b\hatbfx$. The inversion 
time $T_{0}$ is large compared to the inverse Larmor frequency 
$\omega_{0}^{-1}$ 
(viz.\ adiabatic), though small compared to the thermal relaxation time 
$\tau_{th}$ (viz.\ rapid). It provides a highly precise method for inverting 
the qubit Bloch vector $\mathbf{s}=\langle\bfsig\rangle$, although the price 
paid for this precision is an adiabatic inversion rate. We are interested in a 
type of rapid passage in which the control field $\bfF (t)$ as seen in the
detector frame is allowed to twist around in the $x$--$y$ plane with azimuthal 
angle $\phi (t)$ while simultaneously undergoing inversion along the $z$-axis:
\begin{equation}
\label{trpprofile}
\bfF (t) = b\cos\phi (t)\,\hatbfx + b\sin\phi (t)\,\hatbfy + at\,\hatbfz
  \hspace{0.1in} .
\end{equation}
Here $-T_{0}/2\leq t \leq T_{0}/2$ and $\hatbfy =\hatbfz\times\hatbfx$. Note 
that any pair of orthogonal unit vectors in the $x$--$y$ plane can be used
for $\hatbfx$ and $\hatbfy$. Different choices simply alter the value of 
$\phi (t=0)$. As will be explained shortly, interesting physical effects arise 
when the twist profile $\phi (t)$ is chosen appropriately. This type of rapid 
passage is referred to as twisted rapid passage (TRP). The first experimental 
realization of TRP in 1991 by Zwanziger et.~al.\ \cite{zw1} carried out the 
inversion adiabatically with $\phi (t) = Bt^{2}$. Since then, non-adiabatic 
TRP has been studied with polynomial twist profile $\phi (t) = (2/n)B t^{n}$ 
\cite{fg1}, and controllable quantum interference effects were found to arise 
for $n\geq 3$. Zwanziger et.\ al.\ \cite{zw2} implemented non-adiabatic 
polynomial TRP with $n=3,4$ and observed the predicted interference effects. 
In the following subsection we briefly summarize how these quantum 
interferences arise and refer the reader to Ref.~\cite{fg1} for further 
discussion.

\subsection{Controllable Quantum Interference}
\label{sec2.1}
\noindent
In the Zwanziger experiments \cite{zw1,zw2}, a TRP sweep is produced by 
sweeping the detector frequency linearly through resonance at the Larmor 
frequency $\omega_{0}$: $\dot{\phi}_{det}(t) = \omega_{0}+(2at)/ \hbar$. The 
frequency of the rf-field $\dot{\phi}_{rf}$ is also swept through 
resonance in such a way that $\dot{\phi}_{rf}(t) = \dot{\phi}_{det}(t)-
\dot{\phi}(t)$, where $\phi (t)= (2/n)Bt^{n}$ is the TRP twist profile. Thus,
\begin{eqnarray}
\dot{\phi}_{rf}(t) & = & \dot{\phi}_{det}(t)-\dot{\phi}(t)  \nonumber \\
 & = & \omega_{0}+\frac{2at}{\hbar} + \dot{\phi}(t) \hspace{0.1in} .
\label{rescondprelim}
\end{eqnarray}
At resonance $\dot{\phi}_{rf}(t)=\omega_{0}$. Inserting this condition into 
eq.~(\ref{rescondprelim}), it follows that at resonance:
\begin{equation}
\label{rescon}
at -\frac{\hbar}{2}\frac{d\phi}{dt} = 0 \hspace{0.1in} .
\end{equation}
As shown in Ref.~\cite{fg1}, for polynomial twist $\phi (t) = (2/n)Bt^{n}$
with $n\geq 3$, eq.~(\ref{rescon}) has $n-1$ roots, though only the 
real-valued roots correspond to resonance. The various possibilities are 
summarized in Table~\ref{table1}.
\begin{table}[hb]
\tcaption{\label{table1}Classification of regimes under which multiple qubit 
resonances occur for polynomial twist $\phi (t)=(2/n)Bt^{n}$ with $n\geq 3$.}
\centerline{\footnotesize\smalllineskip
\begin{tabular}{lrl}\\
\hline\hline\\
\multicolumn{3}{c}{1.   $\underline{B > 0}$} \\
   (a) $n$ odd:   &   $\:\: 2$ resonances at &
$\:\: t=0$ and $t=\left( a/\hbar B\right)^{\frac{1}{n-2}}$ \\
   (b) $n$ even:  &  $\:\: 3$ resonances at &
$\:\: t=0$ and $t=\pm\left( a/\hbar B\right)^{\frac{1}{n-2}}$ \\
\multicolumn{3}{c}{2.   $\underline{B < 0}$} \\
   (a) $n$ odd:   &   $\:\: 2$ resonances at &
$\:\: t=0$ and $ t=-\left( a/\hbar |B|\right)^{\frac{1}{n-2}}$ \\
   (b) $n$ even:   & $\:\: 1$ resonance at &
$\:\: t=0$ \vspace{0.1in}\\
\hline\hline\\
\end{tabular}}
\end{table}
We see that: (i)~for $B>0$ a qubit always passes through resonance multiple 
times during a \textit{single\/} TRP sweep; (ii)~for $B<0$ multiple resonances 
only occur 
when $n$ is odd; and (iii)~the time separating qubit resonances can be altered 
by variation of the sweep parameters $B$ and $a$. Ref.~\cite{fg1} showed that 
these multiple resonances have a strong influence on the qubit transition
probability. It was shown that qubit transitions could be significantly
enhanced or suppressed by small variation of the sweep parameters, and hence
of the time separating the resonances. Plots of the transition probability
versus time suggested that the multiple resonances were producing quantum
interference effects that could be controlled by variation of the TRP sweep
parameters. In Ref.~\cite{fg2} the qubit transition amplitude was calculated 
to all orders in the non-adiabatic coupling. The result found there can be 
re-expressed as the following diagrammatic series:
\begin{equation}
\label{quantint}
\setlength{\unitlength}{0.05in}
T_{-}(t) = \begin{picture}(10,5)
              \put(10,-1.5){\vector(-1,0){3.25}}
              \put(5,-1.5){\line(1,0){1.75}}
              \put(5,-1.5){\vector(0,1){3.25}}
              \put(5,1.75){\line(0,1){1.75}}
              \put(5,3.5){\vector(-1,0){3.25}}
              \put(0,3.5){\line(1,0){1.75}}
           \end{picture} 
\hspace{0.05in} + 
           \begin{picture}(20,5)
              \put(20,-1.5){\vector(-1,0){3.25}}
              \put(15,-1.5){\line(1,0){1.75}}
              \put(15,-1.5){\vector(0,1){3.25}}
              \put(15,1.75){\line(0,1){1.75}}
              \put(15,3.5){\vector(-1,0){3.25}}
              \put(10,3.5){\line(1,0){1.75}}
              \put(10,3.5){\vector(0,-1){3.25}}
              \put(10,-1.5){\line(0,1){1.75}}
              \put(10,-1.5){\vector(-1,0){3.25}}
              \put(5,-1.5){\line(1,0){1.75}}
              \put(5,-1.5){\vector(0,1){3.25}}
              \put(5,1.75){\line(0,1){1.75}}
              \put(5,3.5){\vector(-1,0){3.25}}
              \put(0,3.5){\line(1,0){1.75}}
           \end{picture} 
\hspace{0.05in} +
           \begin{picture}(30,5)
              \put(30,-1.5){\vector(-1,0){3.25}}
              \put(25,-1.5){\line(1,0){1.75}}
              \put(25,-1.5){\vector(0,1){3.25}}
              \put(25,1.75){\line(0,1){1.75}}
              \put(25,3.5){\vector(-1,0){3.25}}
              \put(20,3.5){\line(1,0){1.75}}
              \put(20,-1.5){\vector(-1,0){3.25}}
              \put(20,3.5){\vector(0,-1){3.25}}
              \put(20,-1.5){\line(0,1){1.75}}
              \put(15,-1.5){\line(1,0){1.75}}
              \put(15,-1.5){\vector(0,1){3.25}}
              \put(15,1.75){\line(0,1){1.75}}
              \put(15,3.5){\vector(-1,0){3.25}}
              \put(10,3.5){\line(1,0){1.75}}
              \put(10,3.5){\vector(0,-1){3.25}}
              \put(10,-1.5){\line(0,1){1.75}}
              \put(10,-1.5){\vector(-1,0){3.25}}
              \put(5,-1.5){\line(1,0){1.75}}
              \put(5,-1.5){\vector(0,1){3.25}}
              \put(5,1.75){\line(0,1){1.75}}
              \put(5,3.5){\vector(-1,0){3.25}}
              \put(0,3.5){\line(1,0){1.75}}
           \end{picture} 
\hspace{0.05in} + \hspace{0.05in} \cdots \hspace{0.25in} .
\end{equation}
Lower (upper) lines correspond to propagation in the negative (positive)
energy level and the vertical lines correspond to transitions between the two 
energy levels. The calculation sums the probability amplitudes for all 
interfering alternatives \cite{f+h} that allow the qubit to end up in the 
positive energy level at time $t$ given that it was initially in the 
negative energy level. As we have seen, varying the TRP sweep parameters 
varies the time separating the resonances. This in turn changes the value of 
each diagram in eq.~(\ref{quantint}), and thus alters the interference between 
alternatives in this quantum superposition. Similar diagrammatic series can be 
worked out for the remaining $3$ combinations of final and intial states. It 
is the sensitivity of the individual alternatives/diagrams to the time 
separation of the resonances that allows TRP to manipulate this quantum 
interference. Zwanziger et.\ al.\ \cite{zw2} observed these interference 
effects in the transition probability using liquid state NMR and found 
quantitative agreement between theory and experiment. It is the link between
the TRP sweep parameters and this quantum interference that we believe makes
it possible for TRP to drive highly accurate single-qubit gates that operate 
non-adiabatically. The results presented in Section~\ref{sec3} for the
different single-qubit gates are found by numerical simulation of the 
one-qubit Schrodinger equation. We next briefly describe how these simulations 
are done~\cite{fg1}.

\subsection{Simulation Protocol}
\label{sec2.2}
\noindent
As is well-known, the Schrodinger dynamics implements a unitary transformation
$U(t,t_{0})$ of the initial quantum state $|\psi (t_{0})\rangle$:
\begin{equation}
|\psi (t)\rangle = U(t,t_{0})\, |\psi (t_{0})\rangle \hspace{0.1in} .
\end{equation}
An $n$-qubit quantum gate implements a fixed unitary transformation $U$
on $n$ qubits. The unitary transformations $U_{H}$,
$U_{P}$, $U_{\pi /8}$, and $U_{NOT}$ carried out by the one-qubit Hadamard, 
phase, $\pi /8$, and NOT gates are, respectively, 
\begin{equation}
\label{Udefs1}
    \hspace{-0.5in}          U_{H} = \frac{1}{\sqrt{2}}
                       \left( \begin{array}{cc}
                               1 & 1 \\
                               1 & -1
                              \end{array}
                       \right)  \hspace{0.25in}
    \hspace{-0.05in}      U_{P} = 
                       \left( \begin{array}{cc}
                               1 & 0 \\
                               0 & i 
                              \end{array}
                       \right)
\end{equation}
 \begin{equation}
\label{Udefs2}
             U_{\pi /8} = 
                        \left( \begin{array}{cc}
                                1 & 0 \\
                                0 & e^{i\pi /4}
                               \end{array}
                        \right) \hspace{0.25in}
             U_{NOT} = 
                        \left( \begin{array}{cc}
                                0 & 1 \\
                                1 & 0 
                               \end{array}
                        \right) \hspace{0.1in} .
\end{equation}
All matrices are in the representation spanned by the computational basis
states $|0\rangle$ and $|1\rangle$ which are chosen to be eigenstates of
$\sigma_{z}$: 
\begin{displaymath}
\sigma_{z}\, |i\rangle = \left( -1\right)^{i}|i\rangle \hspace{0.2in}
   (i=0,1) \hspace{0.1in} .
\end{displaymath}

To determine the dynamical impact of TRP, the $1$-qubit Schrodinger equation 
is simulated numerically in the non-rotating frame in which the Hamiltonian
$H(t)$ is given by eqs.~(\ref{nonrotHam})~and~(\ref{trpprofile}). It is found 
that the numerical stability of the simulation is enhanced if we expand the 
state $|\psi (t)\rangle$ in the instantaneous energy eigenstates $|E_{\pm}(t)
\rangle$ for which $H(t)|E_{\pm}(t)\rangle = E_{\pm}(t)|E_{\pm}(t)\rangle$. 
Because of the direct connection between these states and $H(t)$, they carry 
substantial dynamical information, and a substantial portion of the dynamics 
due to $H(t)$ can be accounted for by choosing this basis. This makes 
the task of determining the remaining dynamics using the Schrodinger equation 
much simpler and the simulation more stable. We thus write: 
\begin{eqnarray}
\label{wavefcn}
\lefteqn{|\psi (t)\rangle  =  S(t)\exp\left[\, -\frac{i}{\hbar}
          \int_{-T_{0}/2}^{t}\, d\theta\left( E_{-} -\hbar\dot{\gamma}_{-}
            \right)\,\right] |E_{-}(t)\rangle } \nonumber \\
 & & \hspace{1.0in} -I(t)\exp\left[\, -\frac{i}{\hbar}\int_{-T_{0}/2}^{t}\, 
  d\theta \left( E_{+}-\hbar\dot{\gamma}_{+}\right)\,\right]|E_{+}(t)\rangle
      \hspace{0.05in} .
\end{eqnarray}
Here $\gamma_{\pm}(t)$ are the adiabatic geometric phases \cite{sha} 
associated with the energy levels $E_{\pm}(t)$, respectively, and
\begin{displaymath}
\dot{\gamma}_{\pm}(t) = i\langle E_{\pm}(t)|\frac{d}{dt}|E_{\pm} (t)\rangle
  \hspace{0.1in} .
\end{displaymath}
Substituting eq.~(\ref{wavefcn}) into the Schrodinger equation leads to the
equations of motion for $S(t)$ and $I(t)$:
\begin{eqnarray}
\label{SIeqmot}
\frac{dS}{dt} & = & -\Gamma^{\ast}(t)\,\exp\left[\, -i\int_{-T_{0}/2}^{t}\,
                     d\theta\,\delta (\theta )\,\right]\, I(t) \nonumber \\
\frac{dI}{dt} & = & \Gamma (t)\,\exp\left[\, i\int_{-T_{0}/2}^{t}\,
                   d\theta\,\delta (\theta )\,\right]\, S(t) \hspace{0.1in} ,
\end{eqnarray}
where
\begin{eqnarray*}
\delta (t) & = & \frac{E_{+}(t)-E_{-}(t)}{\hbar} - \left[\,
   \dot{\gamma}_{+}(t) - \dot{\gamma}_{-}(t)\,\right] \\
\Gamma (t) & = & \langle E_{+}(t)|\frac{d}{dt}|E_{-}(t)\rangle 
                  \hspace{0.1in} ,
\end{eqnarray*}
and $\Gamma^{\ast}(t) = -\langle E_{-}(t)|d/dt|E_{+}(t)\rangle$. The qubit is
initially placed in one of the initial instantaneous energy eigenstates
$|\psi (-T_{0}/2)\rangle = |E_{\pm}(-T_{0}/2)\rangle$ which fixes the
initial condition for $S(t)$ and $I(t)$ through eq.~(\ref{wavefcn}). It proves
useful to recast eqs.~(\ref{SIeqmot}) in dimensionless form. To that end
one introduces the dimensionless time $\tau = (a/b)t$, the dimensionless
inversion rate $\lambda = \hbar |a|/b^{2}$, and the dimensionless twist
strength $\eta_{n} = (\hbar B/a)(b/a)^{n-2}$. The connection between these
dimensionless simulation parameters and the experimental sweep parameters is
given in Section~\ref{sec3}. It is straightforward to
show that the resonances in Table~\ref{table1} occur at \cite{fg1}:
\begin{equation}
\label{zerort}
\tau = 0 \hspace{0.1in} ,
\end{equation}
and
\begin{equation}
\label{nonzerort}
\tau = \left(\mathrm{sgn}\,\eta_{n}\right)^{\frac{1}{(n-2)}}\,
        \left[\, \frac{1}{|\eta_{n}|}\,\right]^{\frac{1}{(n-2)}}
          \hspace{0.1in} ,
\end{equation}
though only the real-valued solutions of 
eqs.~(\ref{zerort}) and (\ref{nonzerort}) correspond to qubit resonances. The 
dimensionless version of eqs.~(\ref{SIeqmot}) are the 
equations that are numerically integrated. The simulations allow us to 
determine the actual unitary transformation $U_{a}$ produced by 
a specific assignment of the TRP sweep parameters $T_{0}$, $a$, $b$, $B$, and 
$n$. Section~\ref{sec2.4} will explain how the sweep parameters are iteratively 
modified so as to make $U_{a}$ approach a target gate $U_{t}$ as closely as 
possible. The iterative procedure searches for a sweep parameter set which 
minimizes (an upper bound for) the error probability $P_{e}$ for $U_{a}$ 
relative to $U_{t}$. We next explain how $P_{e}$ and its upper bound are 
determined.

\subsection{Gate Error Probability}
\label{sec2.3}
\noindent
The following argument is for an $N$-dimensional Hilbert space, though
$N=2$ will be the case of interest in this paper. As in Section~\ref{sec2.2},
let $U_{a}$ denote the actual unitary operation produced by a given set of
TRP sweep parameters and $U_{t}$ a target unitary operation we would like
TRP to approximate as closely as possible. Introducing the operators
$D=U_{a}-U_{t}$ and $P=D^{\dagger}D$, and the normalized state $|\psi\rangle$,
we define $|\psi_{a}\rangle = U_{a}|\psi\rangle$ and $|\psi_{t}\rangle =
U_{t}|\psi\rangle$. Now choose an orthonormal basis $|i\rangle$ ($i=1,\ldots ,
N$) such that $|1\rangle\equiv |\psi_{t}\rangle$ and define the state 
$|\xi_{\psi}\rangle$ via
\begin{eqnarray}
|\psi_{a}\rangle & = & |\psi_{t}\rangle + |\xi_{\psi}\rangle
   \label{eq231} \\
 & = & |1\rangle + |\xi_{\psi}\rangle \hspace{0.1in} . \label{eq232}
\end{eqnarray}
Inserting $|\xi_{\psi}\rangle = \sum_{i=1}^{N}\, e_{i}|i\rangle$ into 
eq.~(\ref{eq232}) gives
\begin{equation}
|\psi_{a}\rangle = \left( 1+e_{1}\right)|1\rangle + \sum_{i\neq 1}\, e_{i}
 |i\rangle  \hspace{0.1in} .
\label{eq233}
\end{equation}
Since $|\psi_{t}\rangle = |1\rangle$ is the target state, it is clear from
eq.~(\ref{eq233}) that the error probability $P_{e}(\psi )$ for $U_{a}$ 
(i.~e.~TRP) is
\begin{eqnarray}
P_{e}(\psi ) & = & \sum_{i\neq 1}\, |e_{i}|^{2} \hspace{0.1in} .
 \label{eq235}
\end{eqnarray}
We define the error probability $P_{e}$ for the TRP gate to be
\begin{equation}
P_{e} \equiv \max_{\scriptstyle |\psi\rangle}\, P_{e}(\psi ) 
\hspace{0.1in} .  \label{eq236}
\end{equation}
From eq.~(\ref{eq231}),
\begin{displaymath}
|\xi_{\psi}\rangle = D|\psi\rangle
\end{displaymath}
and
\begin{eqnarray}
\langle\xi_{\psi}|\xi_{\psi}\rangle & = & \langle\psi |D^{\dagger}D|
     \psi\rangle \nonumber \\
 & = & Tr\rho_{\psi}P \hspace{0.1in} , \label{eq23650}
\end{eqnarray}
where $\rho_{\psi} = |\psi\rangle\langle\psi |$. On the other hand,
\begin{eqnarray}
\langle\xi_{\psi}|\xi_{\psi}\rangle & = & \sum_{i=1}^{N}\, |e_{i}|^{2}
 \nonumber \\
 & = & |e_{1}|^{2} + P_{e}(\psi ) \hspace{0.1in} . \label{eq23675}
\end{eqnarray}
Combining eqs.~(\ref{eq23650}) and (\ref{eq23675}) gives
\begin{eqnarray*}
P_{e}(\psi ) & = & \langle\xi_{\psi}|\xi_{\psi}\rangle - |e_{1}|^{2} \\
 & \leq & \langle\xi_{\psi}|\xi_{\psi}\rangle = Tr\rho_{\psi} P 
 \hspace{0.1in} .
\end{eqnarray*}
Since $P=D^{\dagger}D$ is
Hermitian it can be diagonalized: $P=O^{\dagger}d\, O$ and $d=diag(d_{1},
\ldots , d_{N})$. Thus
\begin{displaymath}
P_{e}(\psi ) \leq Tr\,\overline{\rho}_{\psi}d \hspace{0.1in} ,
\end{displaymath}
where $\overline{\rho}_{\psi} = O\rho_{\psi}O^{\dagger}$. Let $d_{\ast} =
\max (d_{1}, \ldots ,d_{N})$, then direct evaluation of the trace gives
\begin{eqnarray*}
Tr\,\overline{\rho}_{\psi}d & = & \sum_{i=1}^{N} d_{i}
\left(\overline{\rho}_{\psi}\right)_{ii} \\
 & \leq & \sum_{i=1}^{N} d_{\ast}\left(\overline{\rho}_{\psi}\right)_{ii}
  = d_{\ast}\, Tr\,\overline{\rho}_{\psi} = d_{\ast} \hspace{0.1in} ,
\end{eqnarray*}
where we have used that $Tr\,\overline{\rho}_{\psi}=1$. Thus $P_{e}(\psi )
\leq d_{\ast}$ for \textit{all\/} states $|\psi\rangle$. 
From eq.~(\ref{eq236}), it follows that
\begin{equation}
P_{e}\leq d_{\ast} \hspace{0.1in} ,
\label{eq237}
\end{equation}
so that the largest eigenvalue $d_{\ast}$ of $P$ is an upper bound for the 
gate error probability $P_{e}$. Finally, notice that $P=D^{\dagger}D$ is a 
positive operator so that $d_{i}\geq 0$ for $i=1,\ldots , N$. Thus $d_{\ast}
\leq Tr\, P$ and so
\begin{equation}
P_{e}\leq d_{\ast}\leq Tr\, P \hspace{0.1in} .
\label{eq238}
\end{equation}
Although $Tr\, P$ need not be as tight an upper bound on $P_{e}$ as $d_{\ast}$,
it is much easier to calculate and so is more convenient than $d_{\ast}$ for 
use in the sweep optimization procedure to be described next. 

\subsection{Sweep Optimization Procedure}
\label{sec2.4}
\noindent
To find TRP sweep parameters that yield highly accurate non-adiabatic one-qubit 
gates we used the multi-dimensional downhill simplex method \cite{nrp} to
search for sweep parameters that minimize the upper bound $Tr\, P$ for
the gate error probability $P_{e}$. Although we simulated a number of
different types of polynomial twist, all data presented in Section~\ref{sec3}
will be for quartic twist, $\phi_{4}(\tau ) = \left(\eta_{4}/2\lambda \right)
\tau^{4}$, which yielded the best results. The sweep parameters for quartic
twist are ($\lambda$,$\,\eta_{4}$) which can be thought of as specifying
a point in a $2$-dimensional parameter space. For quartic twist, the downhill 
simplex method takes as input $3$ sets of sweep parameters which specify the 
vertices of a simplex in the $2$-dimensional parameter space. The dynamical 
effects of the TRP sweep associated with each vertex is found by numerically 
integrating the one-qubit Schrodinger equation as described in 
Section~\ref{sec2.2}. The output of the integration 
is the unitary operation $U_{a}$ that a particular sweep applies. The desire 
is to iteratively improve $U_{a}$ so that it approximates as closely as 
possible a target unitary operation $U_{t}$. For each $U_{a}$ we determine 
$P=(U_{a}-U_{t})^{\dagger}(U_{a}-U_{t})$ and evaluate $Tr\, P$. The downhill 
simplex method then iteratively alters the simplex (i.~e.\ one or more of its 
vertices) until sweep parameters are found that yield a local minimum of 
$Tr\, P$. Because this minimum is not global, some starting simplexes will 
give deeper minimums than others. Though there was no gaurantee, it was hoped 
that a starting simplex could be found that yielded $Tr\, P< 10^{-4}$. Some 
trial and error in specifying the starting simplex was thus required, though 
for one-qubit gates, the trial and error procedure eventually proved 
successful and we present our results in the following Section.

\section{Simulation Results}
\label{sec3}
\noindent
All results presented below are for quartic twist
\begin{equation}
\phi (\tau ) = \frac{1}{2}\left(\frac{\eta_{4}}{\lambda}\right)\tau^{4}
 \hspace{0.1in} ,
\end{equation}
where $\tau$, $\lambda$, and $\eta_{4}$ are the dimensionless versions of
time $t$, inversion rate $a$, and twist strength $B$ (Section~\ref{sec2.2}).
For convenience, we re-write their definitions here:
\begin{equation}
\tau = \left(\frac{a}{b}\right) t \hspace{0.15in} ; \hspace{0.15in}
 \lambda = \frac{\hbar |a|}{b^{2}} \hspace{0.15in} ; \hspace{0.15in}
  \eta_{4} = \left(\frac{\hbar b^{2}}{a^{3}}\right) B \hspace{0.1in} .
\end{equation}
The parameter $b$ was introduced in eq.~(\ref{trpprofile}) and is the 
rf field amplitude in an NMR realization of TRP \cite{zw2,fg1}.
All simulations were done with $\lambda > 1$ corresponding to non-adiabatic
inversion \cite{zw2,fg1}, and with $\tau_{0}=aT_{0}/b= 80.000$. 

The translation key connecting our dimensionless simulation parameters and the
experimental sweep parameters used in the Zwanziger experiments \cite{zw1,zw2}
was given in the Appendix of Ref.~\cite{fg1}. We re-write the formulas
for quartic twist here for convenience. Note that Zwanziger's symbol $B$
is here replaced by $\mathcal{B}$ to avoid confusion with our use of the
symbol $B$ in this paper to denote the twist strength. The translation
formulas are:
\begin{eqnarray}
\omega_{1} & = & \frac{2b}{\hbar} \\
A  & = & \frac{aT_{0}}{\hbar} \\    
\mathcal{B}  & = & \frac{BT_{0}^{4}}{2} \\
\lambda  & = & \frac{4A}{\omega_{1}^{2}T_{0}} \\
\eta_{4}  & = & \frac{\mathcal{B}\omega_{1}^{2}}{2A^{3}T_{0}} \hspace{0.1in} .
\label{transkey}
\end{eqnarray}
In the experiments of Ref.~\cite{zw2}: $\omega_{1}=393Hz$; $T_{0}= 41.00ms$; 
$A=50\, 000Hz$; and $\mathcal{B}$ was calculated from eq.~(\ref{transkey}) with
$\eta_{4}$ varying over the range $\mathrm{[} 4.50,4.70\mathrm{]}\times 
10^{-4}$.

Note that
$U_{P}$ and $U_{\pi /8}$ (see eqs.~(\ref{Udefs1}) and (\ref{Udefs2})) can
be re-written as
\begin{eqnarray}
U_{P} & = & e^{i\pi /4}\: U_{NOT}\: V_{P} \label{Udefs3}\\
U_{\pi /8} & = & e^{i\pi /8}\: U_{NOT}\: V_{\pi /8} \hspace{0.1in} ,
 \label{Udefs4}
\end{eqnarray}
where
\begin{equation}
\hspace{-0.25in}
V_{P} = \left(  \begin{array}{cc}
                   0 & e^{i\pi /4} \\
                  e^{-i\pi /4} & 0
                \end{array}
        \right)
\label{VPop}
\end{equation}
\begin{equation}
V_{\pi /8} = \left(  \begin{array}{cc}
                   0 & e^{i\pi /8} \\
                  e^{-i\pi /8} & 0
                \end{array}
        \right) \hspace{0.1in} ,
\label{Vpi8}
\end{equation}
and $U_{NOT}$ is given in eq.~(\ref{Udefs2}). As will be seen below, our 
simulations produced $V_{P}$ and $V_{\pi /8}$, from which $U_{P}$ and 
$U_{\pi /8}$ can be constructed using eqs.~(\ref{Udefs3}) and (\ref{Udefs4}),
respectively. TRP is thus used to construct the set of gates $\mathcal{S}_{1} 
= \{ U_{H},V_{P},V_{\pi /8},U_{NOT}\}$ which is universal for one-qubit 
unitary gates. We stress that all gates in this set are produced using a 
non-composite TRP sweep (eq.~(\ref{trpprofile})). The different gates result 
from different choices for the TRP sweep parameters. For each one-qubit gate,
we present our best-case results and show how gate performance is altered by 
small variations in the sweep parameters. 

\subsection*{Hadamard Gate} 

The sweep parameters $\lambda = 5.8511$ and $\eta_{4}=2.9280\times 10^{-4}$
produce the gate $U_{a}$ whose real and imaginary parts are:
\begin{eqnarray}
\hspace{-0.3in} Re(U_{a}) & \hspace{-0.05in} = & \hspace{-0.05in}
   \left( \begin{array}{cc}
                     0.708581 & 0.705629 \\
                     0.705629 & -0.708581
                   \end{array}
            \right)  \\
\hspace{-0.3in} Im(U_{a}) & \hspace{-0.05in} = & \hspace{-0.05in}
   \left( \begin{array}{cc}
                     0.380321\times 10^{-9} & -0.144317\times 10^{-4} \\
                     0.144317\times 10^{-4} & 0.420313\times 10^{-9}
                   \end{array}
            \right) .
\end{eqnarray}
For comparison, the real and imaginary parts of the target Hadamard gate 
$U_{t} = U_{H}$ are:
\begin{eqnarray}
Re(U_{H}) & = & \left( \begin{array}{cc}
                     0.707107 & 0.707107 \\
                     0.707107 & -0.707107
                   \end{array}
            \right) \\
Im(U_{H}) & = & \left( \begin{array}{cc}
                     0 & 0 \\
                     0 & 0
                   \end{array}
            \right) \hspace{0.1in} .
\end{eqnarray}
From $U_{a}$ and $U_{H}$ we find  $Tr\, P=8.82\times 10^{-6}$ and so 
the gate error probability satisfies $P_{e}\leq 8.82\times 10^{-6}$. 
Table~\ref{table2} shows how gate performance varies when the sweep parameters 
are altered slightly. 
\begin{table}[h!]
\begin{center}
\tcaption{\label{table2}Variation of $Tr\, P$ for the Hadamard gate when the
TRP sweep parameters are altered slightly from their best performance values. 
The columns to the left of center have $\eta_{4}=2.9280\times 10^{-4}$ and 
those to the right have $\lambda= 5.8511$\vspace{0.1in}.}
\begin{tabular}{ccc||ccc}\hline
$\eta_{4}$ & $\lambda$ & $Tr\, P$ & 
   $\lambda$ & $\eta_{4}$ & $Tr\, P$ \\ \hline
$2.9280\times 10^{-4}$ & $5.8510$ & $7.22\times 10^{-5}$ &
   $5.8511$ & $2.9279\times 10^{-4}$ & $7.03\times 10^{-4}$ \\
         & $5.8511$ & $8.82\times 10^{-6}$ &
            & $2.9280\times 10^{-4}$ & $8.82\times 10^{-6}$ \\
         & $5.8512$ & $1.84\times 10^{-5}$ &
            & $2.9281\times 10^{-4}$ & $6.14\times 10^{-4}$ \\\hline
\end{tabular}
\end{center}
\end{table}
Of the two sweep parameters, $\eta_{4}$ variation is seen to have the largest 
impact on gate performance. This will turn out to be true for the other 
one-qubit gates as well. Although TRP can produce a Hadamard gate whose error 
probability falls below the accuracy threshold $P_{a}\sim 10^{-4}$, it is
clear from Table~\ref{table2} that the sweep parameters must be controlled to 
$5$ significant figures to achieve this level of performance. See
Section~\ref{sec4} for further discussion this point.

\subsection*{$V_{P}$ Gate} 

As noted above, the target gate here is $V_{P}$, and $U_{P}$ follows from
eq.~(\ref{Udefs3}). The sweep parameters
$\lambda = 5.9750$ and $\eta_{4}=3.8060\times 10^{-4}$ produce the gate 
$U_{a}$:
\begin{eqnarray}
Re(U_{a})  & = & 
   \left( \begin{array}{cc}
                     -0.627432\times 10^{-2} & 0.706181 \\
                     0.706181 & 0.627432\times 10^{-2}
                   \end{array}
            \right) \\
Im(U_{a})  & = & 
   \left( \begin{array}{cc}
                     -0.284521\times 10^{-10} & 0.708004 \\
                     -0.708004 & 0.694222\times 10^{-11}
                   \end{array}
            \right) .
\end{eqnarray} 
From eq.~(\ref{VPop}), the target gate $V_{P}$ is:
\begin{eqnarray}
Re(V_{P}) & = & \left( \begin{array}{cc}
                     0 & 0.707107 \\
                     0.707107 & 0
                   \end{array}
            \right) \\
Im(V_{P}) & = & \left( \begin{array}{cc}
                     0 & 0.707107 \\
                     -0.707107 & 0
                   \end{array}
            \right) \hspace{0.1in} .
\end{eqnarray}
From $U_{a}$ and $V_{P}$ we find $Tr\, P=8.20\times 10^{-5}$ and so 
$P_{e}\leq 8.20\times 10^{-5}$ for this gate. Table~\ref{table3} shows how
$Tr\, P$ varies when $\eta_{4}$ and $\lambda$ are varied slightly. 
\begin{table}[h!]
\begin{center}
\tcaption{\label{table3}Variation of $Tr\, P$ for the $V_{P}$ gate when the
TRP sweep parameters are altered slightly from their best performance values. 
The columns to the left of center have $\eta_{4}=3.8060\times 10^{-4}$ 
and those to the right have $\lambda= 5.9750$\vspace{0.1in}.}
\begin{tabular}{ccc||ccc}\hline
$\eta_{4}$ & $\lambda$ & $Tr\, P$ & 
   $\lambda$ & $\eta_{4}$ & $Tr\, P$ \\ \hline
$3.8060\times 10^{-4}$ & $5.9749$ & $1.56\times 10^{-4}$ &
   $5.9750$ & $3.8059\times 10^{-4}$ & $2.29\times 10^{-3}$ \\
         & $5.9750$ & $8.20\times 10^{-5}$ &
            & $3.8060\times 10^{-4}$ & $8.20\times 10^{-5}$ \\
         & $5.9751$ & $1.43\times 10^{-4}$ &
            & $3.8061\times 10^{-4}$ & $1.88\times 10^{-3}$ \\\hline
\end{tabular}
\end{center}
\end{table}
Again gate performance is most sensitive to variation of $\eta_{4}$, and
the sweep parameters must be controlled to $5$ significant figures for
performance to surpass the accuracy threshold. The latter point is discussed
further in Section~\ref{sec4}. 

\subsection*{$V_{\pi /8}$ Gate} 

From eq.~(\ref{Udefs4}), $U_{\pi /8}$ is found from $V_{\pi /8}$ and $U_{NOT}$.
The target gate this time is $V_{\pi /8}$. For $\lambda = 6.0150$ and 
$\eta_{4}=8.1464\times 10^{-4}$ TRP produced the gate $U_{a}$:
\begin{eqnarray}
Re(U_{a})  & = & 
   \left( \begin{array}{cc}
                     0.101927\times 10^{-2} & 0.925307 \\
                     0.925307 & -0.101927\times 10^{-2}
                   \end{array}
            \right)  \\
Im(U_{a})  & = & 
   \left( \begin{array}{cc}
                     -0.960223\times 10^{-10} & 0.379218 \\
                     -0.379218 & 0.184961\times 10^{-10}
                   \end{array}
            \right) .
\end{eqnarray}
From eq.~(\ref{Vpi8}), the target gate $V_{\pi /8}$ is:
\begin{eqnarray}
Re(V_{\pi /8}) & = & \left( \begin{array}{cc}
                     0 & 0.923880 \\
                     0.923880 & 0
                   \end{array}
            \right) \\
Im(V_{\pi/8}) & = & \left( \begin{array}{cc}
                     0 & 0.382683 \\
                     -0.382683 & 0
                   \end{array}
            \right) \hspace{0.1in} .
\end{eqnarray}
These matrices give $Tr\, P=3.03\times 10^{-5}$ and so for this gate
$P_{e}\leq 3.03\times 10^{-5}$. Table~\ref{table4} shows how gate performance
varies when the sweep parameters are altered slightly. 
\begin{table}[h!]
\begin{center}
\tcaption{\label{table4}Variation of $Tr\, P$ for the $V_{\pi /8}$ gate when 
the TRP sweep parameters are altered slightly from their best performance 
values. The columns to the left of center have $\eta_{4}=8.1464\times 10^{-4}$ 
and those to the right have $\lambda= 6.0150$\vspace{0.1in}.}
\begin{tabular}{ccc||ccc}\hline
$\eta_{4}$ & $\lambda$ & $Tr\, P$ & 
   $\lambda$ & $\eta_{4}$ & $Tr\, P$ \\ \hline
$8.1464\times 10^{-4}$ & $6.0149$ & $1.30\times 10^{-3}$ &
   $6.0150$ & $8.1463\times 10^{-4}$ & $1.77\times 10^{-3}$ \\
         & $6.0150$ & $3.03\times 10^{-5}$ &
            & $8.1464\times 10^{-4}$ & $3.03\times 10^{-5}$ \\
         & $6.0151$ & $2.18\times 10^{-3}$ &
            & $8.1465\times 10^{-4}$ & $2.77\times 10^{-3}$ \\\hline
\end{tabular}
\end{center}
\end{table}
As with the previous two gates, performance is most sensitive to variation of 
$\eta_{4}$, and the sweep parameters must be controllable to $5$ significant
figures (see Section~\ref{sec4}).

\subsection*{NOT Gate} 

Finally, we examine $U_{NOT}$. For $\lambda = 7.3205$ and $\eta_{4}=2.9277
\times 10^{-4}$ TRP produced the gate $U_{a}$:
\begin{eqnarray}
Re(U_{a})  & = & 
   \left( \begin{array}{cc}
                     0.235039\times 10^{-2} & 0.999997 \\
                     0.999997 & -0.235039\times 10^{-2}
                   \end{array}
            \right)  \\
Im(U_{a})  & = & 
   \left( \begin{array}{cc}
                     -0.323648\times 10^{-10} & -0.115151\times 10^{-4} \\
                      0.115150\times 10^{-4} &  0.271006\times 10^{-10}
                   \end{array}
            \right) .
\end{eqnarray}
For comparison, $U_{NOT}$ is (eq.~(\ref{Udefs2})):
\begin{eqnarray}
Re(U_{NOT}) & = & \left( \begin{array}{cc}
                     0 & 1 \\
                     1 & 0
                   \end{array}
            \right) \\
Im(U_{NOT}) & = & \left( \begin{array}{cc}
                     0 & 0 \\
                     0 & 0
                   \end{array}
            \right) \hspace{0.1in} .
\end{eqnarray}
These matrices yield $Tr\, P= 1.10\times 10^{-5}$ and so $P_{e}\leq 1.10\times
10^{-5}$. Table~\ref{table5} shows how $Tr\, P$ varies with small variation
of the sweep parameters. 
\begin{table}[h!]
\begin{center}
\tcaption{\label{table5}Variation of $Tr\, P$ for the NOT gate when 
the TRP sweep parameters are altered slightly from their best performance 
values. The columns to the left of center have $\eta_{4}=2.9277\times 10^{-4}$ 
and those to the right have $\lambda= 7.3205$\vspace{0.1in}.}
\begin{tabular}{ccc||ccc}\hline
$\eta_{4}$ & $\lambda$ & $Tr\, P$ & 
   $\lambda$ & $\eta_{4}$ & $Tr\, P$ \\ \hline
$2.9277\times 10^{-4}$ & $7.3204$ & $1.12\times 10^{-5}$ &
   $7.3205$ & $2.9276\times 10^{-4}$ & $1.23\times 10^{-3}$ \\
         & $7.3205$ & $1.10\times 10^{-5}$ &
            & $2.9277\times 10^{-4}$ & $1.10\times 10^{-5}$ \\
         & $7.3206$ & $1.22\times 10^{-5}$ &
            & $2.9278\times 10^{-4}$ & $1.23\times 10^{-3}$ \\\hline
\end{tabular}
\end{center}
\end{table}
As with the other gates, performance is most sensitive to variation in
$\eta_{4}$, and sweep parameters must be controllable to $5$ significant
figures for the gate error probability $P_{e}$ to fall below the accuracy 
threshold $P_{a}\sim 10^{-4}$ (see Section~\ref{sec4}).

\section{Discussion}
\label{sec4}
\noindent
In this paper we have presented numerical simulation results which suggest 
that TRP sweeps should be capable of producing a set of quantum gates that is 
universal for one-qubit unitary operations. We also showed that sweep 
parameters can be found which the simulations indicate will yield gates that 
operate non-adiabatically and with error probabilities satisfying $P_{e}\leq 
10^{-4}$. To achieve this degree of accuracy, however, the sweep parameters 
must be controllable to high precision ($5$ significant figures). This 
raises the question of whether such precision is possible with 
current technology. In the NMR realization of TRP \cite{zw1,zw2} the inversion 
time $T_{0}$ was of order $10^{-2}\, s$, while the spectrometer waveform 
resolution allowed the rf- and detector-phases to be specified in time steps 
of order $10^{-7}\, s$. Thus $T_{0}$ can be determined to $1$ part in 
$10^{5}$. By using shimming and sample rotation the uncertainty in the 
Larmor frequency $\omega_{0}$ (which is used in eqs.~(\ref{phideq}) and 
(\ref{phiaeq})) can be reduced to $15\, Hz$, while its value is $500MHz$. This 
corresponds to a relative error of $\Delta\omega_{0}/\omega_{0}\sim 10^{-7}$. 
It is also possible to use $\pi$ or $\pi /2$ pulses to calibrate the rf field 
strength $\omega_{1}$ down to a relative error of $\Delta\omega_{1}/\omega_{1}
\sim 10^{-4}$. Thus many of the TRP sweep parameters are already at or near 
the level of precision needed to make high-fidelity one-qubit gates. Still, it 
is clear that further theoretical work is needed to find ways to make the gate 
error probability a more slowly varying function of the TRP sweep parameters. 
Recall that TRP sweeps are non-composite. It is an interesting open question 
whether composite sweeps that interlay TRP with different types of pulses can 
lead to more robust gate performance. We intend to examine this question in 
our next set of simulations. Having discussed current challenges, it is worth 
stressing that these sweeps show genuine potential for producing high-fidelity 
non-adiabatic one-qubit gates. Further work to try to develop this potential 
seems warranted. Although other approaches exist for making one-qubit gates 
(e.~g.~Ref.~\cite{epr}), in a field faced with as many technical challenges 
as quantum computing, it is advantageous to have multiple ways to accomplish
important tasks. It is hoped that with further development TRP gates may 
provide an approach to making high-fidelity non-adiabatic quantum gates.
TRP sweeps also provide a concrete example of how quantum effects can be used 
to enhance our control of a quantum system. Further study of these sweeps also 
seems worthwhile as a question of basic physics.  

\subsection*{Atomic Physics}

The following scenario is inspired by the NMR realization of TRP
\cite{zw1,zw2}. Consider electric dipole transitions between a pair of
atomic energy eigenstates $|+\rangle$ and $|-\rangle$ of the Hamiltonian
$H_{a}$ with respective energies $E_{\pm}=\pm\epsilon_{0}/2$. Transition 
between these two states is caused by an applied electric field 
$\mathbf{E}_{a}(t)=2E_{1}\cos\phi_{a}(t)\,\mathbf{x}$ which couples to the
atom's electric dipole moment $\mathbf{d}=e\,\mathbf{r}$. In the lab frame,
the two-level Hamiltonian $\mathcal{H}(t)$ in the rotating wave approximation
is \cite{a+e}:
\begin{displaymath}
\mathcal{H}(t) = -\frac{\hbar\omega_{0}}{2}\,\sigma_{z} + 
                  \frac{\hbar\omega_{1}}{2}\left[\,\cos\phi_{a}(t)\,\sigma_{x}
                   +\sin\phi_{a}(t)\,\sigma_{y}\,\right] \hspace{0.1in} ,
\end{displaymath}
where $\hbar\omega_{0}=\epsilon_{0}$ and $\hbar\omega_{1}=d_{x}E_{1}$.
Transformation to the detector frame \cite{zw1,sut} is done using the unitary
operator $U(t)=\exp [-(i/2)\phi_{det}(t)\sigma_{z}]$ so that $\mathcal{H}
\rightarrow\overline{\mathcal{H}}$: 
\begin{eqnarray}
\overline{\mathcal{H}}(t)  & = & 
 \frac{\hbar}{2}\left(\dot{\phi}_{det}-\omega_{0}\right)\sigma_{z}
  +\frac{\hbar\omega_{1}}{2}\left[
           \cos (\phi_{a}-\phi_{det})\sigma_{x}+\sin (\phi_{a}-\phi_{det})
            \sigma_{y}\right] \vspace{0.20in}\nonumber \\
 & = & at\sigma_{z} + b\cos\phi_{n}(t)\sigma_{x}+b
   \sin\phi_{n}(t)\sigma_{y}
   \hspace{0.1in} , \label{calH}
\end{eqnarray}
where
\begin{eqnarray}
at & = & \frac{\hbar}{2}\left(\,\dot{\phi}_{det}-\omega_{0}\,\right) 
 \label{ateq}\\
b & = & \frac{\hbar\omega_{1}}{2} \\
\phi_{n}(t) & = & \phi_{a} - \phi_{det} \hspace{0.1in} ,\label{phineq}
\end{eqnarray}
and $\phi_{n}(t) = (2/n)Bt^{n}$ is the twist profile for polynomial twist.
Eq.~(\ref{calH}) gives $\overline{\mathcal{H}}(t) = \mbox{\boldmath $\sigma$}
\cdot \mathbf{F}(t)$, where $\mathbf{F}(t)$ is the control field for TRP
appearing in eq.~(\ref{trpprofile}). Integrating eq.~(\ref{ateq}) gives
$\phi_{det}(t)$ which can then be inserted into eq.~(\ref{phineq}) so that
\begin{eqnarray}
\phi_{det}(t) & = & \frac{at^{2}}{\hbar} + \omega_{0}t \label{phideq}\\
\phi_{a}(t) & = & \frac{at^{2}}{\hbar} + \omega_{0}t + \frac{2}{n}Bt^{n}
 \hspace{0.1in} . \label{phiaeq}
\end{eqnarray}
We see that programming the generator that produces $\mathbf{E}_{a}(t)$ so 
that the phase $\phi_{a}(t)$ is given by eq.~(\ref{phiaeq}) causes a TRP sweep 
to be applied to the atom in the detector frame. Note that, to insure the 
two-level approximation is valid, the frequencies $\dot{\phi}_{n}(t)$ swept 
through by the TRP sweep should not include the resonance frequency of any 
other pair of atomic energy levels since this would drive unwanted dynamics 
not included in $\mathcal{H}(t)$. 

\subsection*{Previous Work} 

Recently, Morton et.\ al.\ \cite{epr} showed how to use composite pulses to
produce high fidelity single-qubit operations in electron paramagnetic
resonance. The composite pulses are based on the BB1 corrective sequence
\cite{bb1}. Along with observation of non-decay of Rabi oscillations and 
suppression of secondary Fourier components in the spin echo decay envelope,
they compared an improved Carr-Purcell pulse sequence (in which BB1 composite
$\pi$-pulses replace ordinary $\pi$-pulses) with the Carr-Purcell-Meiboom-Gill
sequence. From the decay of the echo produced by the improved Carr-Purcell
sequence they inferred a fidelity for the BB1 $\pi$-pulses of $\mathcal{F}=
0.9999$. The authors noted that this fidelity is ultimately limited by pulse 
phase errors.  

The fidelity in Ref.~\cite{epr} is $\mathcal{F}=(1/2)\mathrm{Re}\left[ \, Tr
\left( U_{a}^{\dagger}U_{t}\right)\right]$. It is possible to relate our 
$Tr\, P$ upper bound on $P_{e}$ to this fidelity. Recalling that $P=\left( 
U_{a}-U_{t}\right)^{\dagger}\left( U_{a}-U_{t}\right)$, we have
\begin{eqnarray*}
Tr\, P & = & Tr\left( 2 - \left[ U_{a}^{\dagger}U_{t}+U_{t}^{\dagger}U_{a}
           \right]\right) \\
 & = & 4 - 2 \,\mathrm{Re}\left[\, Tr\left( U_{a}^{\dagger}U_{t}\right)\right] 
  \\
 & = & 4\left( 1-\mathcal{F}\right) \hspace{0.1in} ,
\end{eqnarray*}
and so
\begin{equation}
\label{fidlty}
\mathcal{F} = 1 - \frac{1}{4}\, Tr\, P \hspace{0.1in} .
\end{equation}
Using the results from Section~\ref{sec3} for $Tr\, P$ in eq.~(\ref{fidlty}),
we can determine the fidelity for the TRP gates: 
\begin{eqnarray}
\mathcal{F}_{H} & = & 0.9999\, 98 \label{fida}\\
\mathcal{F}_{V_{P}} & = & 0.9999\, 80 \\
\mathcal{F}_{V_{\pi /8}} & = & 0.9999\, 92 \\
\mathcal{F}_{NOT} & = & 0.9999\, 97 \label{fidb} \hspace{0.1in} . 
\end{eqnarray}

\subsection*{Future Work} 
\begin{description}
\item[(a)] We are currently exploring whether TRP can be used to make 
a two-qubit gate that will complete the one-qubit gates considered here to 
give a set that: (i) is universal for quantum computation; and (ii) has all 
gates operating non-adiabatically with fidelities that yield $P_{e}<P_{a}$. 
A progress report on this work will be given elsewhere.
\item[(b)] Development of an approximate analytical approach to TRP would be 
very useful. We are not aware of any general tractable analytical approach to 
non-adiabatic rapid passage that could be used to find good starting simplexes 
for the sweep optimization procedure. It is because of this that we followed
the numerical approach described above.
\item[(c)] Constructing a theory for the optimum twist profile $\phi (t)$ 
for a given quantum gate would also be a valuable contribution. To date, 
quartic twist has worked best, though we do not presently have arguments 
explaining why it will produce better gates than the other examples of TRP 
that we have considered, or whether some other profile will work even better.
\item[(d)] It would be especially interesting if the simulation results 
presented above could be tested experimentally. One possibility might be 
to use state tomography to measure the output density matrix 
$\rho_{exp}=U_{a}|\psi_{0}\rangle\langle\psi_{0}|\, U^{\dagger}_{a}$ 
resulting from an initial state $|\psi_{0}\rangle$, for each of the TRP
generated gates $U_{a}$ presented in Section~\ref{sec3}. Associated with each 
sweep is a target gate $U_{t}$ and a corresponding target density matrix 
$\rho_{t}=U_{t}|\psi_{0}\rangle\langle\psi_{0}|U_{t}^{\dagger}$. Having 
measured $\rho_{exp}$, evaluate the fidelity $\mathcal{F}(\rho_{exp}, 
\rho_{t})$ \cite{n+c}:
\begin{equation}
\mathcal{F}(\rho_{exp},\rho_{t})  =  
   Tr\,\sqrt{(\rho_{exp})^{1/2}\;\rho_{t}\; (\rho_{exp})^{1/2}} 
            \hspace{0.1in}. \label{fidc}
\end{equation}
Although this fidelity differs from the one considered in Ref.~\cite{epr}, one 
might naively anticipate that they are of comparable size. If so, then the 
experimentally determined fidelities should be close to the fidelities given 
in eqs.~(\ref{fida})--(\ref{fidb}).
\item[(e)] The simulation results presented in this paper are for an isolated 
qubit interacting with a noiseless TRP sweep. Although this scenario might
appear idealized, it seemed sensible to see what kind of performance was 
possible using TRP under the best possible conditions. One important extension 
would be to allow the TRP sweeps to include a noise component. To the extent 
that this noise leads to dephasing, TRP gate performance is expected to 
deteriorate once the dephasing time is of order the TRP inversion time 
$T_{0}$. Under these conditions, the qubit dynamics begins to loose its 
temporal phase coherence, and the quantum interference between alternatives 
begins to disappear. It would be worthwhile to consider simple noise models 
to study the sensitivity of TRP gate performance to parameters such as noise 
power and noise correlation time. A study along these lines was done for the 
quantum adiabatic search algorithm in Ref.~\cite{fg3}. Phase decoherence 
resulting from interaction of the target qubit with environmental qubits is 
another source of concern. As with the case of noise above, should the 
decoherence time be of order $T_{0}$, TRP gate performance is expected to 
suffer. Follow-up work that sheds light on how this performance cross-over 
occurs would be valuable. 
\end{description}
\nonumsection{Acknowledgments}
\noindent
M. Hoover was supported by the Illinois Louis Stokes Alliance for Minority 
Participation Bridge to the Doctorate Fellowship, and F. Gaitan thanks
T. Howell III for continued support.

\nonumsection{References}

\end{document}